# Computer Program Decomposition and Dynamic/Behavioral Modeling

**Sabah Al-Fedaghi**

*sabah.alfedaghi@ku.edu.kw*

Computer Engineering Department, Kuwait University, Kuwait

**Summary**

Decomposition—statically dividing a program into multiple units—is a common programming technique for realizing parallelism and refining programs. The decomposition of a sequential program into components is tedious, due to the limitations of program analysis and because sequential programs frequently employ inherently sequential algorithms. This paper contributes to this area of study by proposing a diagrammatic methodology to decompose a sequential program. The methodology involves visualizing the program in terms of a conceptual model called the thinging machine (TM) model. The TM diagram-based model establishes three levels of representation: (1) a static description; (2) a dynamic representation; and (3) a behavioral model. The decomposition is performed in the last phase of modeling, according to the streams of events. This method is contrasted with formal decomposition specifications and compared with the typical decomposition of a C++ program. The results point to the viability of using TM for decomposing programs.

## 1. Introduction

Computer science involves analyzing computational *artifacts* and the methods involved in their design, specification, programming, verification, implementation, and testing. The "abstract nature of computer programs and the resulting complexity of implemented artifacts" have raised many questions and research issues; in particular, "a program can be taken as an abstract thing, [or] it may also be cashed out as a sequence of physical operations" [1]. For example, the assignment statement A: = 13 + 74 is interpreted as an abstract statement and also as a physical memory of location A receiving the value of physically computing 13 times 74 [2].

A software program describes a sequence of instructions, mostly imperative instructions that are communicated through the memory [3]. Thus, "a program is 'in essence' a sequence of instructions" [4]. A sequential program is a series of operations on a set of variables, with each operation being completed before the next one begins [5]. According to Allen [5], "The static representation of the program imposes an ordering on all operations to which the execution must adhere. Every instance of the program running with a particular input yields the same sequence of operations. The sequential execution is predictable, because of the ability to anticipate what a program will do." Sequential programming is an important field of computer science because "it still seems to be the basis of programming, both in practice and in teaching" [4]. Arguably, the sequential execution model is "the most successful abstraction in the field of computer science, serving as the basis for most existing software" [5].

### 1.1 Decomposition for Parallelism

However, the technology paradigm has switched to parallelism. This development is described by Allen [5] as follows. The sequential execution model is amenable to high-performance execution, which has tracked the exponential transistor scaling of Moore's law [6]. Technological forces have limited further increases in processor performance [7]. Therefore, computer architects have utilized the exponentially increasing number of transistors at their disposal to integrate multiple processor cores onto a single chip [8]. Multicore processors offer benefits such as increased performance, improved response time, and decreased power consumption. However, "To leverage these benefits, software must be capable of dividing its constituent computations among the cores of a multicore processor to achieve parallel execution" [5]. To exploit the potential of multicore processors, programmers must decompose programs in a manner suitable for parallel execution.

### 1.2 Decomposition for Program Refinement

One of Abrial's [9] main concerns is building models of programs that are "quite different from the program itself [since] it is far easier to reason about the model than about the program." A program's model, "although not executable, allows us to clearly identify the properties of the future system and to prove that they will be present in it" [9]. We refine a model to decompose it, and we decompose it to further refine it more freely [9].

Decomposition is the process of systematically splitting a model into component models to reduce the complexity and thus of refining each component model independently of the others [9]. The component models can be combined again to form a single model. A program is obtained at the final stage of a sequence that consists of building increasingly accurate models of the program. The program is also decomposed into smaller ones to enhance its readability [9].

### 1.3 About This Paper

This paper is about program decomposition, which is tedious for a sequential program [10]. According to Chen [3], "it is not trivial to rewrite sequential programs with parallelism, nor to write parallel ones from scratch" [3]. There is no general way to execute sequential programs in parallel [5]. Automatic parallelization is ineffective due to the limitations of program analysis and because sequential programs frequently employ inherently sequential algorithms [11] (see [5]). On the other hand, according to Abrial [9], decomposition is done in a "very systematic fashion." Abrial [9] adopted classical set-theoretic notations in his refinement of program models.

The paper proposes a diagrammatic method to address the problem of *how to decompose a sequential program* for both purposes This is accomplished by visualizing the program as a conceptual model called a thinging machine (TM) model. A conceptual model can support communication, learning, and analysis about relevant aspects of the underlying domain and "can serve as a vehicle for reasoning and problem solving, and for acquiring new knowledge" [12]. The TM is a diagram-based model that establishes three levels of representation: (1) a static structural description, which is constructed on the basis of the flow of things in five generic operations (activities, i.e., create, process, release, transfer, and receive); (2) a dynamic representation, which identifies hierarchies of events based on the five generic events; and (3) a behavioral representation according to the chronology of events. The next section presents the background of the TM through a brief review and with a new example.

## 2. Thinging Model Theory

This section briefly reviews TM modeling to provide a basis for applying TM to analyze how to decompose programs. More elaborate discussions of TM's philosophical foundations can be found in [13-20].

### 2.1 Basics of the Thinging Machine

Typically, ontology requires classifications, such as a functional classification of human bodily functions into mental, sensory, speech, respiratory, and digestive functions and so on [21]. Yet, even with the impressive progress in developing ontologies of things (i.e., entities or objects), the ontology of processes (TM machines) has not made similar advances [21]. The TM ontology is *an* ontology of perdurants (processes or events) as opposed to an ontology of endurants (objects) [12]. The TM is a conceptual model that is concerned only with modeling a view of the domain according to a given application. This is in contrast to a design model, which translates a view to a suitable implementation according to the underlying implementation environment [12].

TM is based on a one category of entities [22] called thimacs (*thi*ngs/*ma*chines). The thimac is simultaneously an "object" (called a *thing*) and a "process" (called a *machine*)—thus, the name thimac. The thimac notion is not new. In physics, subatomic entities must be regarded as both particles and waves to fully describe and explain observed phenomena [23]. According to Sfard [24], abstract notions can be conceived of in two fundamentally different ways: structurally, as objects/things (static constructs), and operationally, as processes. Thus, distinguishing between form and content and between process and object is popular, but, "like waves and particles, they have to be united in order to appreciate light" [25]. TM adopts this notion of duality in conceptual modeling and generalizes it beyond mathematics.

The term "thing" relies more on Heidegger's [26] notion of "things" than it does on the classical notion of objects. According to Heidegger [26], a thing is self-sustained, self-supporting, or independent—it is something that stands on its own. More importantly, it is that which can be spoken about, "that which can be talked about [or] that which is named" [27]. "Talking about" a thing denotes the thing being modeled in terms of being created, processed (change), released, transferred, and/or received. According to Johnson [27], "there is *no* thing that we cannot speak about." In Heidegger's [26] words, a thing "things"; that is, it ties its constituents together in the *same way that a bridge unifies environmental aspects* (e.g., a stream, its banks, and the surrounding landscape). In our TM ontology of dual being, the thing's machine (the machine side of the thing) "machines"; that is, it operates on (other) things by creating, processing, releasing, transferring, and/or receiving them.

The term "machine" refers to a special abstract machine called a TM (see Fig. 1). A central premise underlying the TM is that its performance is limited to five generic operations: creating, processing (changing), releasing, transferring, and receiving. A thimac has dual being as a thing and as a machine. A thing is created, processed, released, transferred, and/or received. A machine creates, processes, releases, transfers, and/or receives things. We will alternate among the terms "thimac," "thing," and "machine" according to the context.

The five TM operations (also called stages) form the foundation for thimacs. Among the five stages, flow (a solid arrow in Fig. 1) signifies conceptual movement from one machine to another or among a machine's stages. The TM's stages can be described as follows.
- *Arrival*: A thing reaches a new machine.
- *Acceptance*: A thing is permitted to enter the machine. If the machine always accepts arriving things, then arrival and acceptance can be combined into the "receive" stage. For simplicity, this paper's examples presume the existence of a receive stage.
- *Processing* (change): A thing undergoes a transformation that changes it without creating a new thing.
- *Release*: A thing is marked as ready to be transferred outside of the machine.
- *Transference*: A thing is transported somewhere outside of the machine.
- *Creation*: A new thing is born (created) within a machine. A machine creates in the sense that it finds or originates a thing; it brings a thing into the system and then becomes aware of it.

Creation can designate "bringing into existence" within the system because what exists is what is found. Additionally, creation does not necessarily mean existence in the sense of being alive. Creation in a TM also means appearance within the system. Here, appearance is not limited to form or solidity but also applies to any sense of the system's awareness of the new thing. Even nominals (which have no existence except as names) may be things that appear in the system model.

In addition, the TM model includes memory and triggering (represented as dashed arrows), or relations among the processes' stages (machines). For example, the process in Fig. 2 triggers the creation of a new thing.

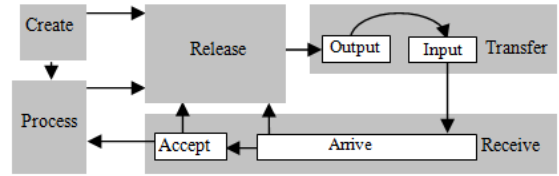
Fig. 1. A thinging machine.

2.2 Example

As mentioned in the introduction, programs have a dual nature: they have an abstract guise as well as a physical one [28]. The assignment A: = 13 + 74 is interpreted as an abstract statement and also as a physical memory of location A receiving the value of physically computing 13 times 74 [2].

A TM applies duality regardless of the domain. Consider Fig. 2, which represents A: = 13 + 74 as a thimac in the abstract domain. It has dual being as a machine and a thing. In the abstract, the (abstract) machine processes 13 and 74 to create the thing 87. The creation on the machine side corresponds to manifestation on the thing side. Fig. 3 illustrates A: = 13 + 74 in the physical domain. It has dual being as a machine and a thing. In reality, the machine ALU processes 13 and 74 to create the data 87. In TM modeling, we show only the machine side of the thimac.

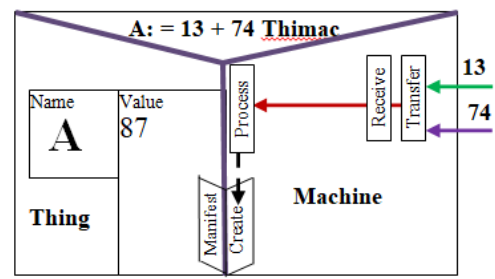
Fig. 2. In the abstract, the *thi* A exists through creation by its *mac*.

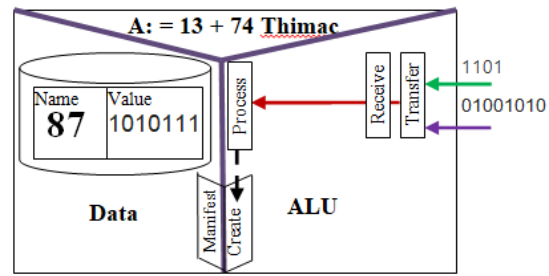
Fig. 3. Physically, the *thi* exists through creation by its *mac*.

## 3. Decomposing a Program

Programs are made up of assignment statements, conjoined together by a number of operators, specifically sequential compositions, conditionals, and loops. Abrial [9] gives an example of a sequential program, which is shown in Fig. 4. The expression *swap (g, k + 1, j + 1)* represents the swapping of the values $g(k + 1)$ and $g(j + 1)$ in the array *g*. According to Abrial [9], decomposition has four components, which we list later.

### 3.1 TM Static Model

Fig. 5 shows the static model of the program of Fig. 4, which is needed to decompose the program using TM.
- In the diagram of Fig. 5, the value of *j* (circle 1 in the figure) flows to be compared with *m* (2). If $j \neq m$ then $g(j + 1)$ as follows.

The variable *j* and the constant *1* are triggered (4 and 5, respectively) to flow, where they are processed to generate $j + 1$ (6). The result $j + 1$ flows to *g* (7 and 8). In *g*, $j + 1$ is processed (9) to generate $g(j + 1)$ (10), which flows to storage (11) (Being stored is a prerequisite to swap $g(j+1)$ with $g(k+1)$ later). The result $g(j + 1)$ is also processed (compared) with *x* (12). Accordingly,

```
while j ≠ m do
    if g(j + 1) > x then
        j := j + 1
    elsif k = j then
        k, j := k + 1, j + 1
    else
        k, j, g := k + 1, j + 1, swap (g, k + 1, j + 1)
    end
end ;
p := k
```

Fig. 4. Sample program (adopted from Abrial [9]).

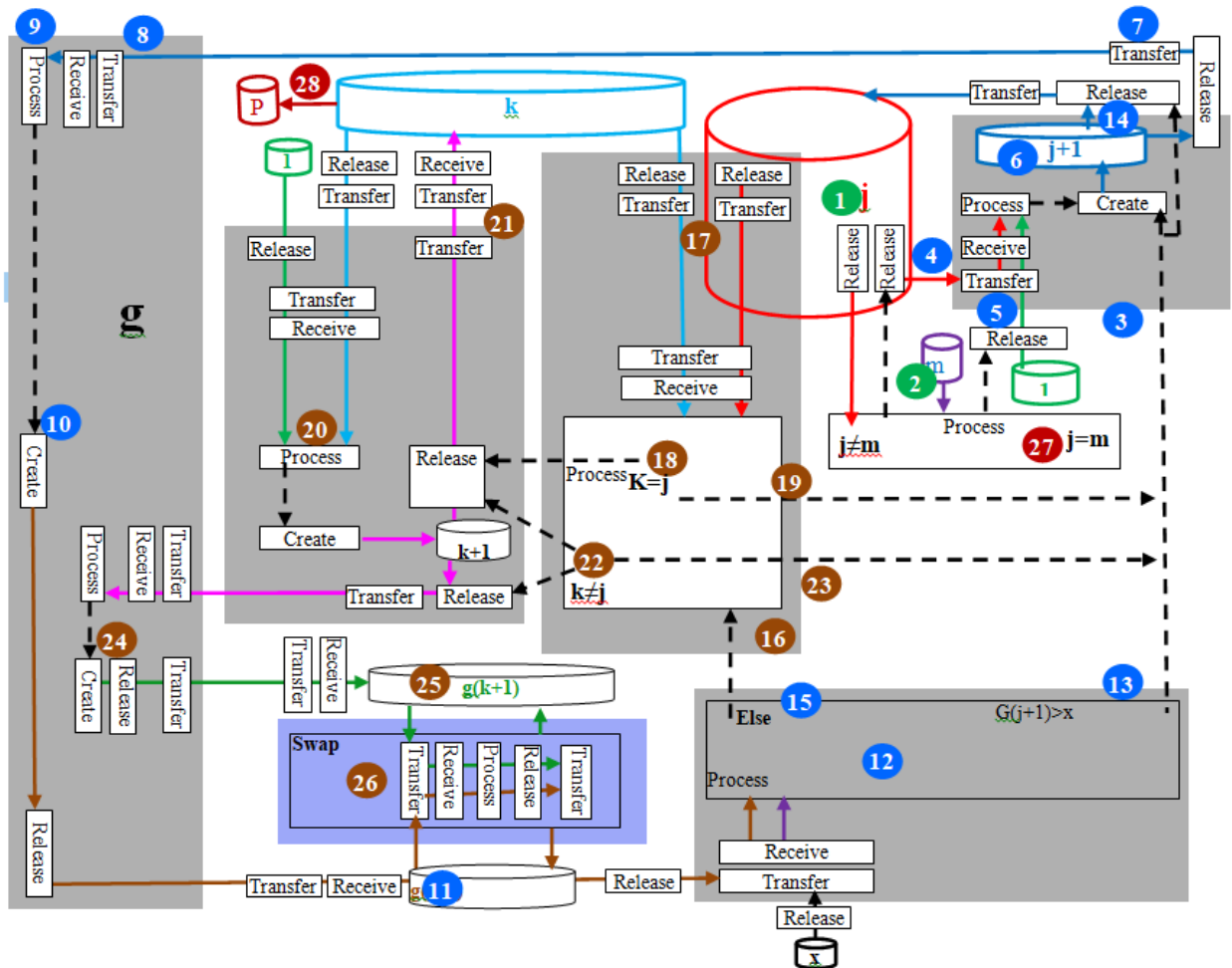

Fig. 5. The TM static model of the program in Fig. 4.

- If $g(j + 1) > x$ (13), then $j + 1$ replaces $j$ (14).
- Else (15) $j$ and $k$ are processed (16).
- Processing $j$ and $k$ involves the flows of $j$ and $k$ (17). Accordingly,
  - $j+1$ replaces $j$ (19) and $k$ is incremented by 1 (20) and $k+1$ replaces $k$ (21).
  - If $k \neq j$ (22), then $k + 1$ flows to $g$ (23), where it is processed to generate $g(k + 1)$ (24), which flows to be stored (25).
- Then, $g(k + 1)$ (11) and $g(j + 1)$ are swapped (26).
- Returning to comparing $j$ and $m$ (27 – red circle), determine that $j = n$; then, $k$ flows to $p$ (28). For simplicity, we did not draw the triggering arrow from 27 to 28.

3.2 TM dynamic Model

The dynamic model involves identifying events. An event in a TM is a thimac with a time subthimac.

For example, Fig. 6 shows the TM representation of the event *Processing* $g(k + 1)$ *and* x. Fig. 6 (dark box) shows the region where the event occurs. For simplicity, we represent each event by its region, assuming that no two events have the same region and time.

Fig. 7 shows the dynamic model with the selected events listed as follows.

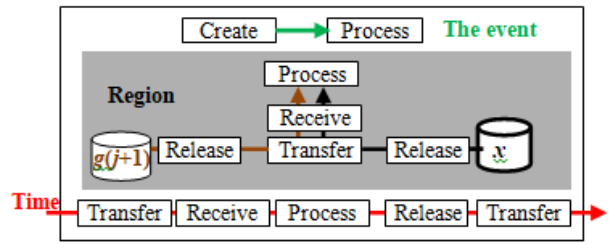

Fig. 6. The event *Processing (comparing)* $g(j + 1)$ *and* x.

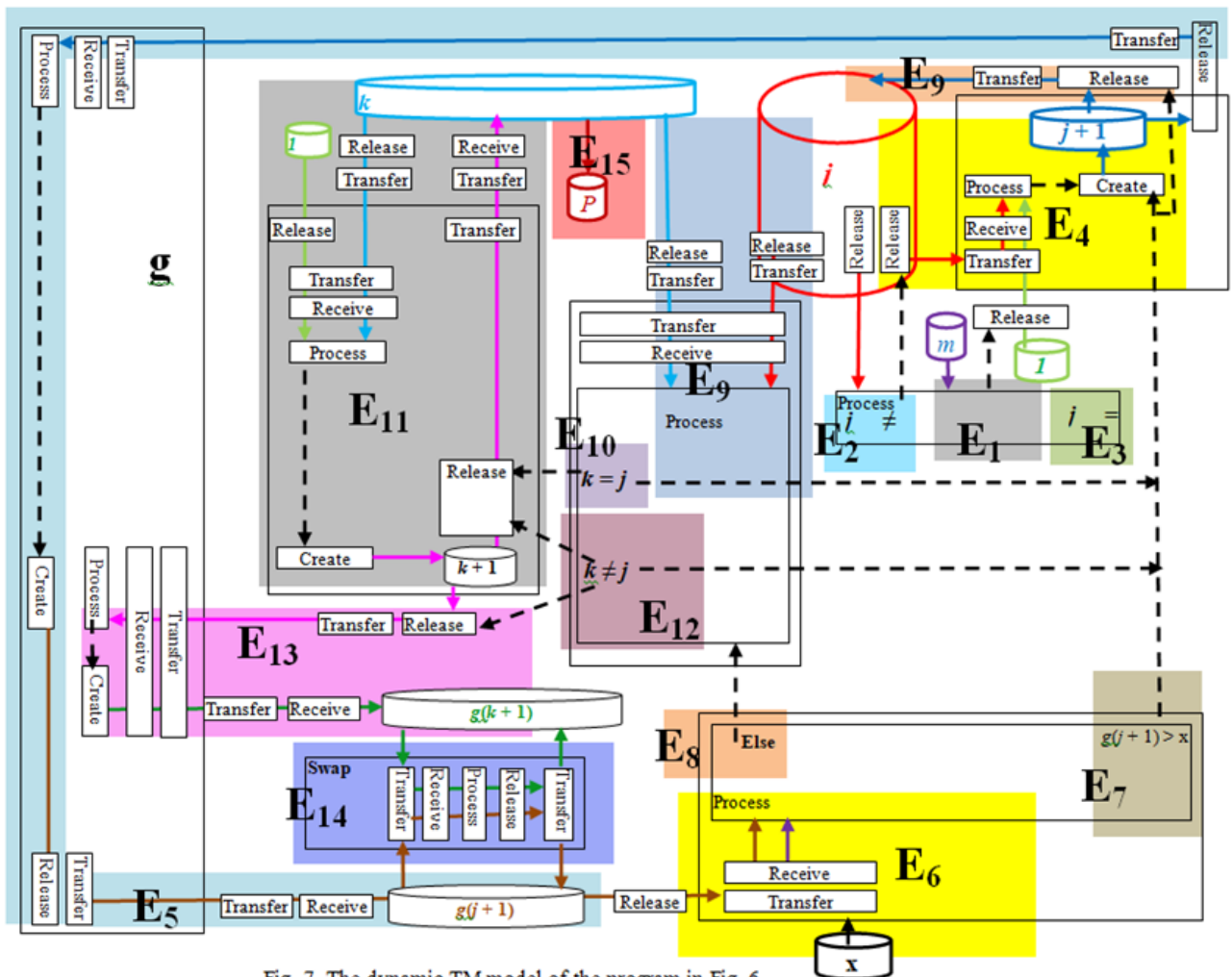

Fig. 7. The dynamic TM model of the program in Fig. 6.

Event 1 ($E_1$): Processing $j$ and $m$
Event 2 ($E_2$): $j \neq m$
Event 3 ($E_3$): $j = m$
Event 4 ($E_4$): Calculating $j + 1$
Event 5 ($E_5$): Storing $j + 1$ in $j$
Event 6 ($E_6$): Calculating and storing $g(j + 1)$
Event 7 ($E_7$): Processing $g(j + 1)$ and $x$
Event 8 ($E_8$): $g(j + 1) \Rightarrow x$
Event 9 ($E_9$): Else
Event 10 ($E_{10}$): Processing $k$ and $j$
Event 11 ($E_{11}$): Incrementing $k$ by 1 and storing in $k$
Event 12 ($E_{12}$): $k \neq j$
Event 13 ($E_{13}$): Calculating and storing $g(k + 1)$
Event 14 ($E_{14}$): Swapping $g(k + 1)$ with $g(k + 1)$
Event 15 ($E_{15}$): Flowing of $k$ to $p$

Fig. 8 shows the behavioral model in terms of the chronology of the events $E_1$ through $E_{15}$. As indicated in the figure, it is not difficult to identify the given program's decomposition from the diagram. According to Abrial [9], "This decomposition has been done in a very systematic fashion." The TM presents an alternative way of approaching program decomposition.

## 4. Bank Transaction System

Allen [5] presents a typical bank-transaction-processing program used to execute bank transactions, which a practical system would run in real-time and which would require a reactive concurrent solution. The bank account operations are too fine-grained to be parallelized on current multiprocessors. The code for a bank account class is listed in Fig. 9. Fig. 10 shows a multithreaded version of the bank-transaction-processing example.

The program's decomposition divides the transaction input into chunks and assigns each chunk of work to a thread. This figure is intended to contrast such a programming-based decomposition with its corresponding representation in TM.

```
1 // Read bank transactions one at a time,
2 // until there are no more transactions.
3   for (trans_t* trans = get_next_trans(); trans != NULL;
4       trans = get_next_trans()) {
5     account_t* account = trans->account;
6
7     if (trans->type == DEPOSIT)
8       account->deposit(trans->amount);
9
10    else if (trans->type == WITHDRAW)
11      account->withdraw(trans->amount);
12
13    else if (trans->type == BALANCE)
...
```
Fig. 9. Bank-transaction processing (partially from [5]).

```
...
8 // Break the transactions into chunks of equal size,
9 // and assign each chunk to a thread.
10 int trans_per_thread = transactions.size() / NUM_THREADS;
11 thread_info_t thread_info[NUM_THREADS];
12 thread_t tid[NUM_THREADS];
13 for (int i = 0; i < NUM_THREADS; ++i) {
14   thread_info[i].begin = trans_per_thread * i;
15   thread_info[i].end = thread_info[i].begin + trans_per_thread;
...
```
Fig. 10. Threads of transaction (partially from [5]).

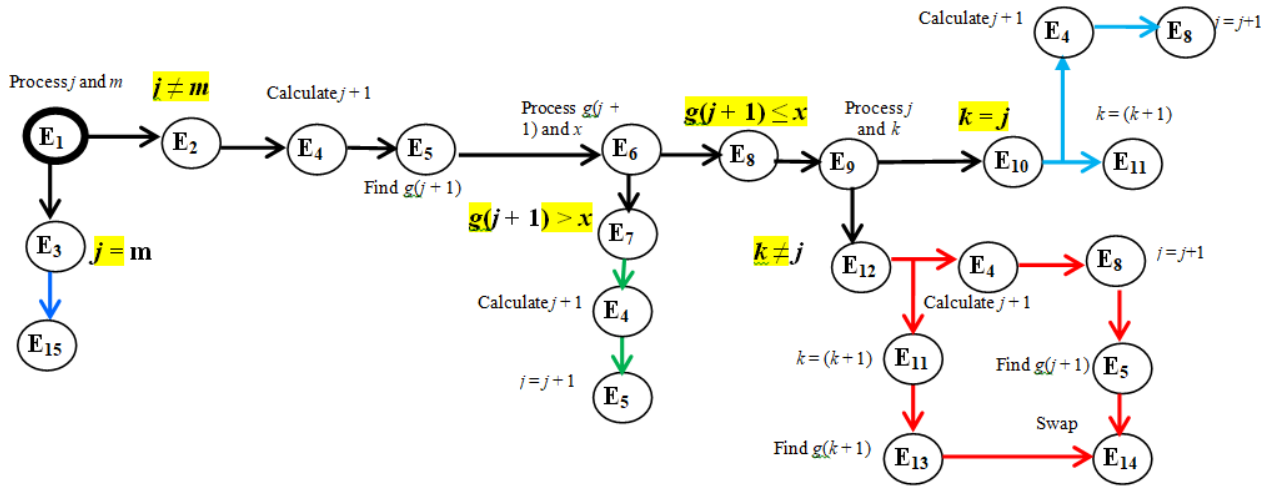

Fig. 8. The behavioral model.

Fig. 11 represents the bank-transaction-processing system diagrammatically as a TM model that can be described as follows.
- A bank transaction is created (circle 1) that includes data about the account number, type of transaction (deposit or withdraw), and amount. The transaction flows to the system (2), where it is processed (3).
- The fields of the transactions are extracted (4-6). Note that extracting here means the arrival (transfer and receive), i.e., the field's appearance in the system. The extracted account number is sent (8) to the database system, where it is processed (9) along with the file (10) to extract the relevant record (11; the record in the file that corresponds to the given account number). This record is sent to the system (12).
- In the system, the record is processed to extract the account number (13) and balance (14). This balance and the amount of the transaction (5) are processed (15) according to the type of transaction (4). A new balance is generated (16) according to whether the type of transaction is a deposit or a withdrawal (17).

A new record is constructed (18) from the account number (13) and the new balance (17). The new record flows to the database system (19). The database system processes (20) the record and the old file to replace the old record with a new record, thus creating a new file (21). Last, an acknowledgement that the transaction has been competed is sent (22).

Figure 12 shows the dynamic model of the bank-transaction system. It includes 17 events; A to Q. Fig. 13 shows the behavioral model of the bank transaction system. It is easy to see different possible compositions for parallel execution. The diagrammatic representation of different streams of events presents an alternative to the method of analyzing a program's text to identify different "chunks" and assigning chunks to threats (see Fig. 10).

Accordingly, we simplified the behavioral model as shown in Fig. 14, where consecutive events are merged as one event. Consequently, it is possible to achieve parallelism with five programs. As an example, and with an assumption of equal time among the five slots or levels, Fig. 15 shows an execution of five programs.

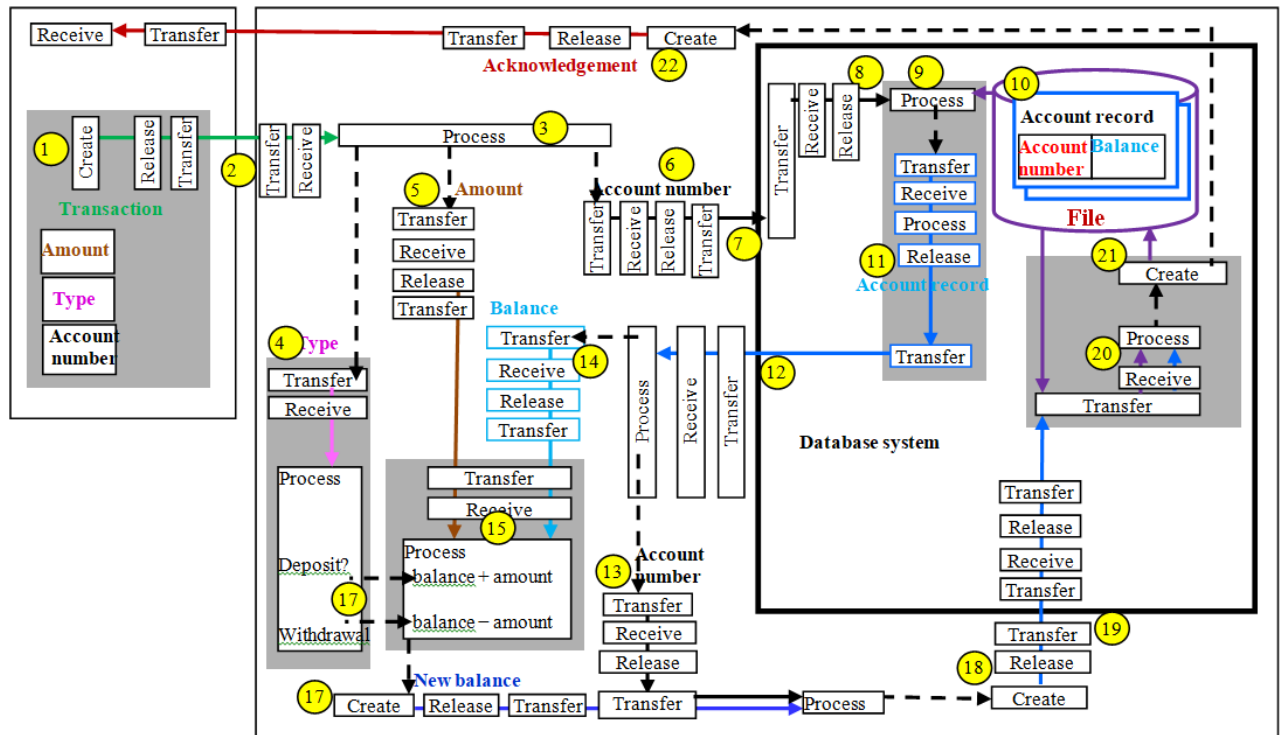

Fig. 11. The static TM model of the bank-transaction process.

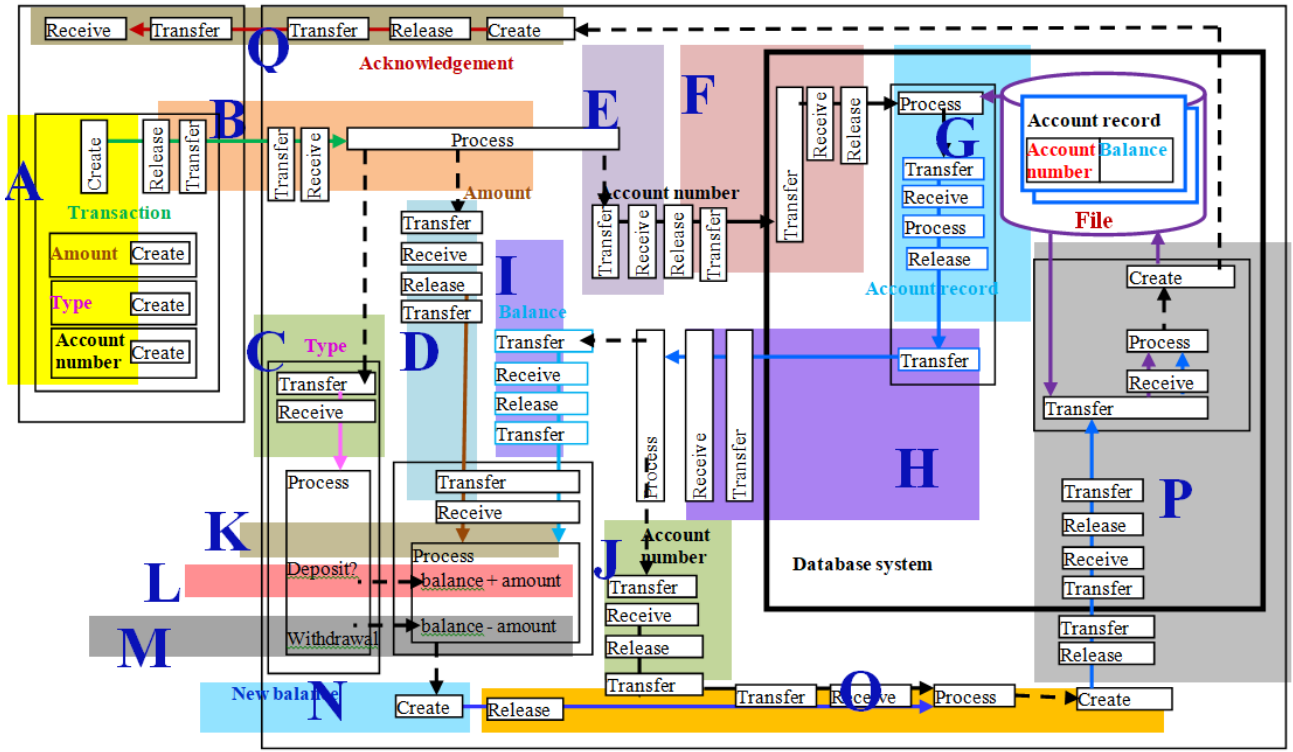

Fig. 12. The dynamic model of the bank-transaction process.

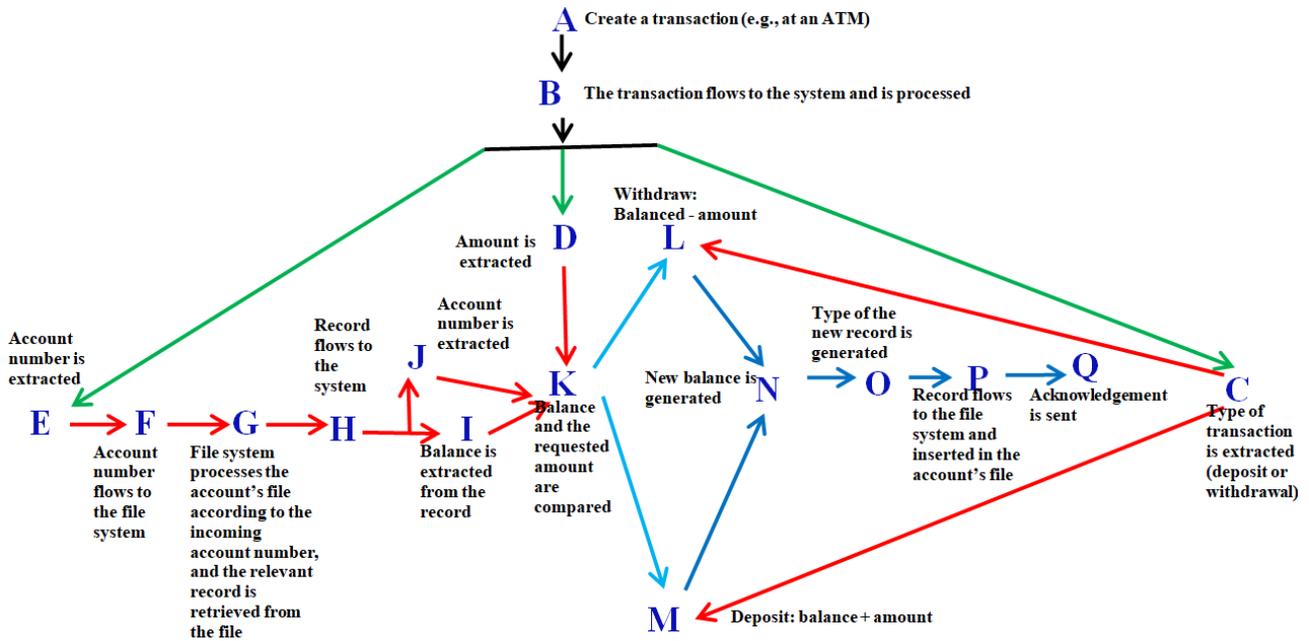

Fig. 13. The behavioral TM model of the bank transaction system.

## 5. Algorithm Decomposition

According to Abrial [9], "The initial model of a program describes the properties that the program must fulfil. It does not describe the algorithm contained in the program, but rather the way by which we can eventually judge that the final program is correct." Abrial [9] modeled the algorithm for a binary search in a sorted array $f$ in a non-decreasing way. First, Abrial [9] presented a mathematical declaration, e.g., axm0_1: $n \in N$, axm0_2: $f \in 1 \dots n \rightarrow N$, axm0_3: $v \in ran(f)$, and thm0_1: $n \geq 1$. Then. Abrial developed three refinements to produce the algorithm.

These refinements involve generating four decompositions by incrementing and decrementing the value of r and comparing $f(r)$ with $v$, assuming that $r$ is an index of $f$ and $v$ is the search value. Fig. 16 shows the increment portion (of $r$), and Fig. 17 shows the final binary search algorithm.

By applying this approach to the TM modeling, it is clear that the model involves, besides the initial condition,
- Incrementing $r$ when $v$ is in the upper half of the segment being searched,
- Decrementing $r$ when $v$ is in the upper half of the segment being searched, and,
- Comparing $f(r)$ with $v$.

By using $p$ and $q$ as pointers to the bottom and top of the currently searched segment of the array, respectively, we can produce the three decompositions shown in Fig. 18. From these initial three compositions and considering the initial conditions segment, we can merge the four segments to produce the TM representation of the binary search shown in Fig. 19.

In Fig. 19, first, $p$ is initialized to 1 (circle 1), $q$ is initialized to $n$ (2), and $r$ is initialized to $(r + 1 + q) / 2$ (3). Next, $r$ is released (4) to be processed with the array $f$ (5), to produce $f(r)$ (6). The array element $f(r)$ and $v$ (7) are compared (8). According to this comparison,
- If $(r) = v$ (9), then $v$ is found and $r$ is the output.
- If $f(r) < v$ (10), then this triggers the dark thimac pointed to by the dashed arrow (11). Note that we can make the arrow point to Process in the dark box, meaning that this box is to be activated. The value of $r$ is retried (11), incremented by 1 (12), and put in $p$ (14). Additionally, $r + 1$ and $q$ are processed (15) to generate $(r + 1 + q) / 2$ (16) and stored as a new value of $r$.

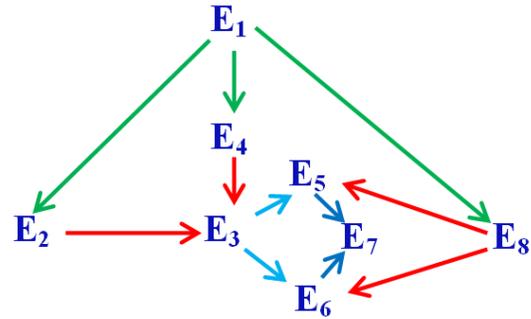

Fig. 14. Simplification of the behavioral model by merging consecutive events.

| Program 1 | Program 1 | Program 1 | Program 1 | Program 1 |
|---|---|---|---|---|
| $E_1$ | $E_2, E_4,$ and $E_8$ | $E_3$ | $E_5$ or $E_6$ | $E_7$ |
| | Program 2 | Program 2 | Program 2 | Program 2 |
| | $E_1$ | $E_2, E_4,$ and $E_8$ | $E_3$ | $E_5$ or $E_6$ |
| | | Program 3 | Program 3 | Program 3 |
| | | $E_1$ | $E_2, E_4,$ and $E_8$ | $E_3$ |
| | | | Program 4 | Program 4 |
| | | | $E_1$ | $E_2, E_4,$ and $E_8$ |
| | | | | Program 5 |
| | | | | $E_1$ |

Fig. 15. Sample parallel execution of five programs.

```
inc
  when
    f(r) < v
  then
    p := r + 1
    r := (r + 1 + q) / 2
  end
```

Fig. 16. The increment decomposition (from [9]).

```
bin_search_program
  p, q, r := 1, n, (1 + n)/2;
  while f(r) = v do
    if f(r) < v then
      p, r := r + 1, (r + 1 + q) / 2
    else q, r := r − 1, (p + r − 1) / 2
  end
end
```

Fig. 17. The binary search program (from [9]).

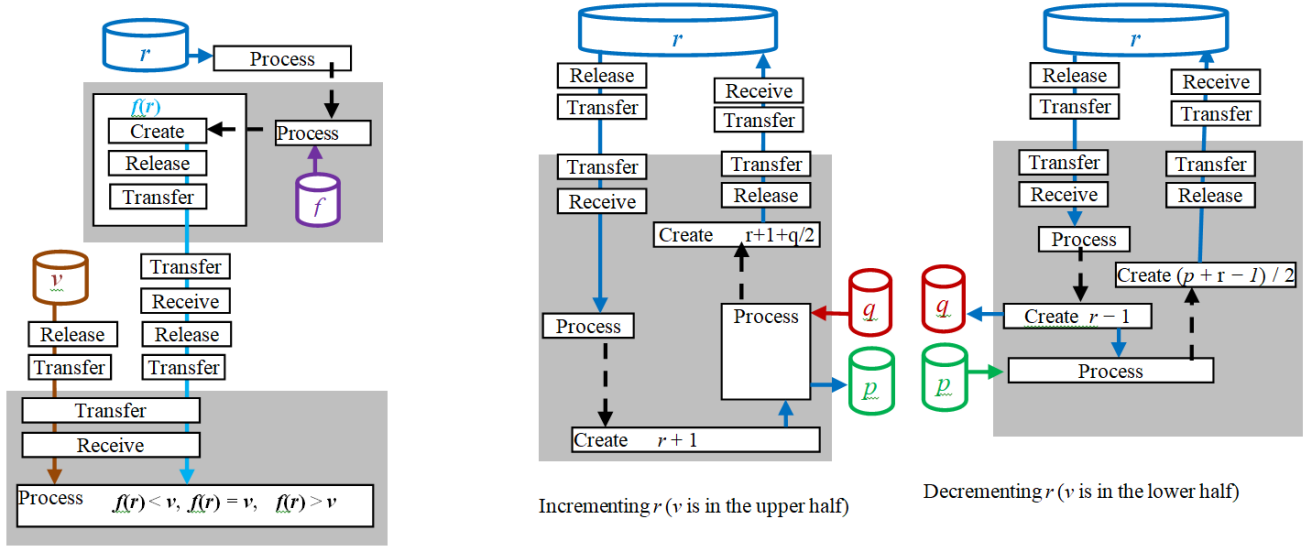

Fig. 18. Three decompositions of the binary search.

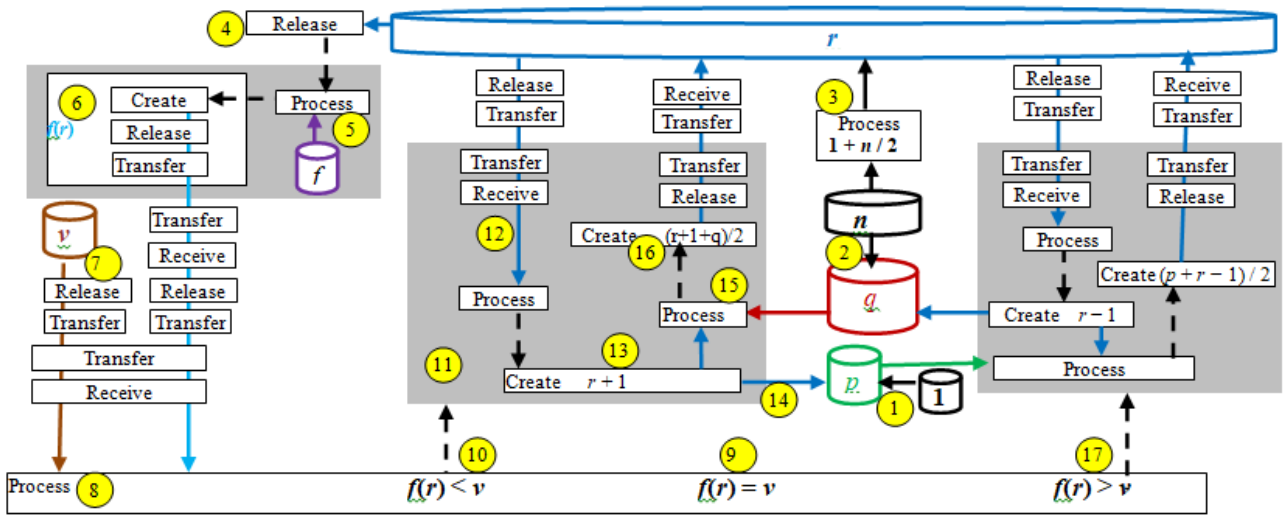

Fig. 19. The TM model of the binary search.

- If $f(r) > v$ (17) then a similar process to the previous point is performed.

Fig. 20 shows the dynamic model for the binary search, and Fig. 21 shows the behavioral model. Note that the three decompositions shown in Fig. 18 are represented as consecutive events (dotted ellipses). Accordingly, we can say that a single composition in a system is a sequence of events (maybe one event) in its behavioral model.

### Conclusion

This paper has demonstrated the applicability and usefulness of the TM model in decomposing programs for parallelism and design refinement. As demonstrated by the examples, TM models can be used as visualization tools for analyzing programs. Future research should use TM modeling in different areas of system decompositions.

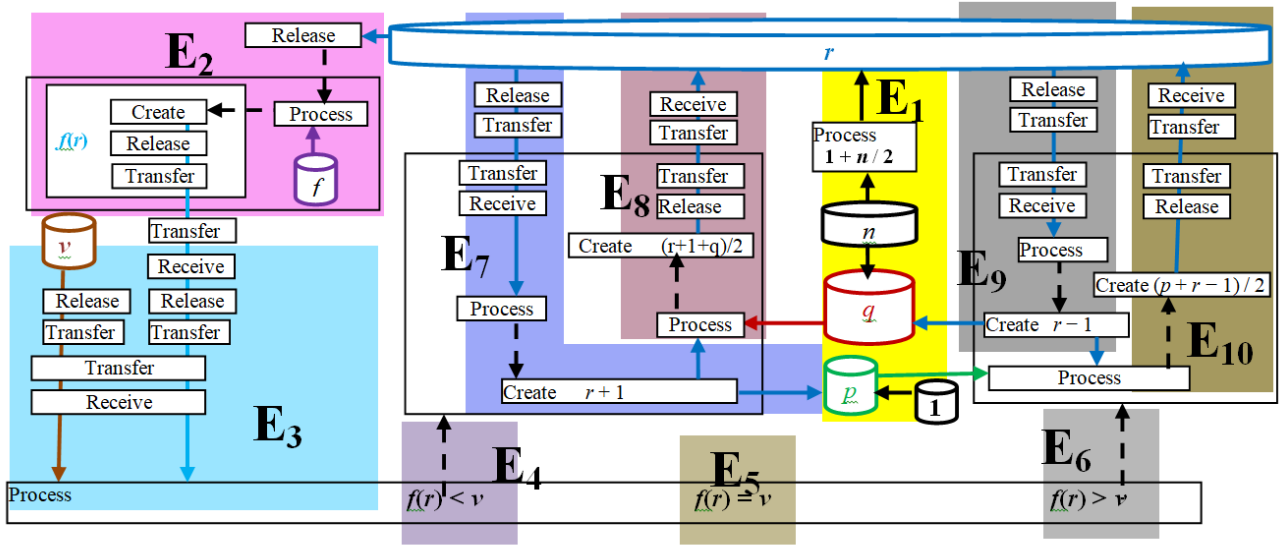

Fig. 20. The dynamic model of the binary search.

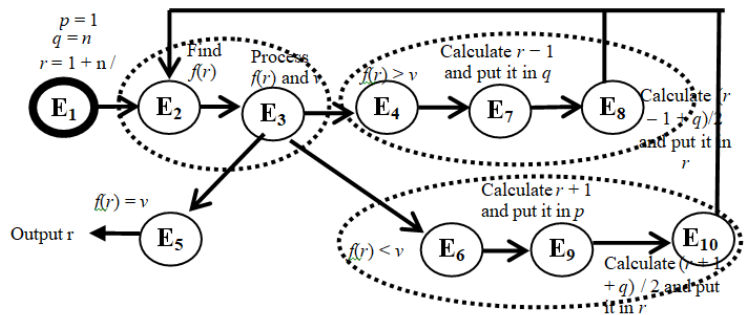

Fig. 21. The behavioral model of the binary search.